\documentclass[b5paper,twoside,reqno]{bjp}
\usepackage{cite}
\usepackage{latexsym}
\usepackage{graphicx}
\usepackage{amssymb,amsmath,amsfonts,amsbsy, amsthm, xspace, latexsym, amscd, enumitem}
\usepackage[utf8]{inputenc}
\usepackage[T1]{fontenc}
\usepackage[english]{babel}
\usepackage{times}
\usepackage{url,bm}
\renewcommand\congress{\small\sf Section: Theoretical Physics}
\usepackage{amsmath,amsfonts,amssymb}

\usepackage{amssymb,amscd,amsmath,amsxtra}

\newcommand{\ov}{\over}
\newcommand{\nn}{\nonumber}

\newcommand{\bea}{\begin{eqnarray}}
\newcommand{\eea}{\end{eqnarray}}

\newcommand{\be}{\begin{equation}}
\newcommand{\ee}{\end{equation}}

\def\a{\alpha}
\def\b{\beta}
\def\g{\gamma}

\def\D{\Delta}

\def\e{\epsilon}

\newcommand{\p}{\partial}

              \def\CC{{\cal C}}
              
\def\CG{{\cal G}}              
              \def\CL{{\cal L}}

\def\a{\alpha}
\def\b{\beta}
\def\g{\gamma}

\def\e{\epsilon}

\def\l{\lambda}

\def\G{\Gamma}
\def\D{\Delta}

\def\ov{\over}

\def\tphi{\tilde\phi}

\def\hepth#1{{arXiv:hep-th/}#1}

\def\np#1#2#3{{Nucl. Phys.} {\bf B#1} (#2) #3}
\def\pl#1#2#3{{Phys. Lett. }{\bf B#1} (#2) #3}

\def\ijmp#1#2#3{{Int. J. Mod. Phys.} {\bf #1} (#2) #3}

\makeindex

\begin{document}

\sloppy \raggedbottom

\title{On the RG Flow in the Two-Dimensional Coset Models}


\begin{start}
\author{Marian Stanishkov}{1}

\address{Institute for Nuclear Research and Nuclear Energy,\\ Bulgarian Academy of Sciences, Sofia 1784, Bulgaria}{1}

\begin{Abstract}
We consider a RG flow in a general $\hat{su}(2)$ coset model perturbed by the least relevant field. The perturbing field as well as some particular fields of dimension close to one are constructed recursively in terms of lower level fields. Using this construction we obtain the structure constants and the four-point correlation functions in the leading order. This allows us to compute the mixing coefficients among the fields in the UV and the IR theory.
\end{Abstract}


\end{start}

In this paper we consider the general $\hat{su}(2)$ coset model $M(k,l)$ \cite{gko} perturbed by the least relevant operator. Recently, there was some interest in the calculations of the renormalization group (RG) properties of such theories, like $N=0$ (Virasoro) \cite{pogt} and its $N=1$ supersymmetric extension \cite{as}. They are just particular cases of $M(k,l)$. It is known \cite{myo} that there exists an infrared fixed point of the renormalization group flow of this theory which coincides with the model $M(k-l,l)$.  Here we are interested in the mixing of certain fields under the corresponding RG flow. It is known that the mixing coefficients coincide for $l=1$ (Virasoro) and $l=2$ (superconformal) theories up to the second order of the perturbation theory \cite{as}. We will show that this is the case in the general theory, i.e. they do not depend on $l$ and are finite up to the second order.
Calculations up to the second order is always a challenge even in two dimensions. The problem is that one needs in addition to the structure constants also the corresponding four-point functions which are not known exactly. Fortunately, it turns out that we need the value of these functions up to the zero order of the small parameter $\epsilon$.
Basic ingredients for the computation of the correlation functions in two dimensions are the conformal blocks. They are quite complicated objects and a close form is not known. We find it convenient, following \cite{myo} , to use the construction presented in \cite{myt} . Namely, we define the perturbing field and the other fields in consideration recursively as a product of lower level fields. Then the corresponding structure constants and four-point functions at some level $l$, governing the perturbation expansion, can be obtained recursively from those of the lower levels or finally from the Virasoro minimal models themselves by certain projected tensor product.

Consider a two-dimensional CFT $M(k,l)$ based on the coset:
\be\nn
{\hat{su}(2)_k\times \hat{su}(2)_l\over \hat{su}(2)_{k+l}},
\ee
where $k$ and $l$ integers, we assume $k>l$. It is written in terms of $\hat{su}(2)_k$ WZNW models with current $J^a$, $k$ is the level. The latter are CFT's with a stress tensor expressed through the currents by the Sugawara construction, the central charge is $c_k={3k\ov k+2}$. The energy momentum tensor of the coset is then $T=T_k+T_l-T_{k+l}$ and:
\be\nn
c={3kl(k+l+4)\over (k+2)(l+2)(k+l+2)}={3l\over l+2}\left(1-{2(l+2)\over (k+2)(k+l+2)}\right).
\ee
The dimensions of the primary fields $\phi_{m,n}(l,p)$ of the "minimal models" ($m,n$ are integers) are computed in \cite{kmq}:
\bea\label{dmn}
\D_{m,n}(l,p) &=&{((p+l)m-p n)^2-l^2\over 4lp(p+l)}+{s(l-s)\over 2l(l+2)},\\
\nn  &=&|m-n|( mod (l)),\hskip1cm 0\le s\le l,\\
\nn &1&\le m\le p-1, \hskip1cm 1\le n\le p+l-1
\eea
where we introduced ${\bf p=k+2}$ (note that we inverted $k$ and $l$  in the definition of the fields).

In this paper we will use a description of the theory
$M(k,l)$ presented in \cite{myt}. It was shown there that this theory is not
independent but can be built out of products  of theories of lower
levels. Schematically this can be written as a recursion:
\be\label{proj}
M(1,l-1)\times M(k,l)={\bf P}(M(k,1)\times M(k+1,l-1))
\ee
where ${\bf P}$ in the RHS is a specific projection. It allows the
multiplication of fields of the same internal indices and describes
primary and descendent fields.

In the following we will be interested in the CFT $M(k,l)$ perturbed by the least
relevant field.  The theory is described by the Lagrangian:
$$
\CL(x)=\CL_0(x)+\l \tilde\phi(x)
$$
where $\CL_0(x)$ describes the theory $M(k,l)$ itself. We define the field $\tilde\phi=\tilde\phi_{1,3}$ in terms of lower level fields:
\be\label{field}
\tilde\phi_{1,3}(l,p)=a(l,p)\phi_{1,1}(1,p)\tilde\phi_{1,3}(l-1,p+1)+b(l,p)\phi_{1,3}(1,p)\phi_{3,3}(l-1,p+1).
\ee
Here the field $\phi_{3,3}(l,p)$ is just a primary field form (\ref{dmn}). The dimension of the field (\ref{field}) is:
\be\label{delt}
\D=\D_{1,3}+{l\over l+2}=1-{2\over p+l}=1-\e.
\ee
In this paper we consider the case $p\rightarrow\infty$ and
assume that $\e={2\over p+l}\ll 1$ is a small parameter. The coefficients $a(l,p)$ and $b(l,p)$ as well as the structure constants of the fields involved in the construction (\ref{field}) can be found by demanding the closure of the fusion rules \cite{myo}.

The mixing of the fields along the RG flow is connected to the two-point function. Up to the second order of the perturbation theory it is given by:
\bea\nn
<\phi_1(x)\phi_2(0)>&=&<\phi_1(x)\phi_2(0)>_0-\l\int <\phi_1(x)\phi_2(0)\tilde\phi(y)>_0 d^2y+\\
\nn &+&{\l^2\ov 2}\int <\phi_1(x)\phi_2(0)\tilde\phi(x_1)\tilde\phi(x_2)>_0 d^2x_1 d^2x_2 +\ldots
\eea
where $\phi_1$, $\phi_2$ can be arbitrary fields of dimensions $\D_1$, $\D_2$. The first order corrections are expressed through the structure constants which we will present later. Let us focus here on the second order. One can use the conformal transformation properties of the fields to bring the double integral to the form:
\bea\label{bint}
&\int& <\phi_1(x)\phi_2(0)\tilde\phi(x_1)\tilde\phi(x_2)>_0 d^2x_1d^2x_2 =\\
\nn &=&(x\bar x)^{2-\D_1-\D_2-2\D}\int I(x_1) <\tilde\phi(x_1)\phi_1(1)\phi_2(0)\tilde\phi(\infty)>_0 d^2x_1
\eea
where
\be\nn
I(x)=\int |y|^{2(a-1)}|1-y|^{2(b-1)}|x-y|^{2c} d^2y
\ee
and $a=2\e+\D_2-\D_1$,$b=2\e+\D_1-\D_2$, $c=-2\e$. It is well known that the integral for $I(x)$ can be expressed in terms of hypergeometric functions whose behaviour around the points $0$, $1$ and $\infty$ is well known. It is clear that the integral (\ref{bint}) is singular. The regularization procedure consists in introducing a parameter $r$ in addition to the usual cut-off $r_0$. They define rings where the integral is convergent. Near the singular points we use the OPE.

Let us consider the correlation function that enters the integral (\ref{bint}). The basic ingredients for the computation of the four-point correlation functions are the conformal blocks. According to the construction (\ref{proj}) any field $\phi_{m,n}(l,p)$ (or its descendent) can be expressed recursively as a product of lower level fields. Therefore the corresponding conformal blocks will be a product of lower level conformal blocks. Due to the RHS of (\ref{proj}) only certain products of conformal blocks will survive the projection ${\bf P}$. The conformal block is a chiral object, i.e. it depends only on the chiral coordinate $x$. It can be expanded as
\be\label{exp}
F(x)=x^{\D_{rs}-\D_{i_1j_1}-\D_{i_2j_2}}\sum_{N=0}^\infty x^N F_N
\ee
where $N$ is called level (not to be confused with the level $l$ of $M(k,l)$) and we omitted the indexes. In order to preserve the projection ${\bf P}$ in the intermediate channel, we allow only products of conformal blocks of the form:
\bea\label{cbprod}
&<&\phi_{i_1,j_1}(x)\phi_{i_2,j_2}(0)|_{r,t}\phi_{i_3,j_3}(1)\phi_{i_4,j_4}(\infty)>_1\times\\
\nn &\times&<\phi_{k_1,l_1}(x)\phi_{k_2,l_2}(0)|_{t,s}\phi_{k_3,l_3}(1)\phi_{k_4,l_4}(\infty)>_{1-1}\times\\
\nn &\times&\sqrt{\CC_{(i_1j_1)(i_2j_2)}^{rt}\CC_{(i_3j_3)(i_4j_4)}^{rt}\CC_{(k_1l_1)(k_2l_2)}^{ts}\CC_{(k_3l_3)(k_4l_4)}^{ts}}.
\eea
Namely, only products of conformal blocks that involve the same internal indexes in the cross channel are allowed. Note that we included explicitly the corresponding structure constants. This is needed because they give different relative contribution on the subsequent levels in the expansion (\ref{exp}). The overall constant will define the actual structure constant. Also, as explained in \cite{myo}, we take square roots of the structure constants because our considerations are chiral, i.e. depend only on the chiral coordinate $x$.
The conformal blocks are in general quite complicated objects.
Fortunately, in view of the renormalization scheme and the
regularization of the integrals, we need to compute them here only
up to the zero-th order in $\e$. This simplifies significantly the
problem. Once the conformal blocks are known, the correlation function of spinless fields for our $M(k,l)$ models is written as:
\be\nn
\sum_{r,s} C_{rs}|F(r,s)|^2
\ee
where the range of $(r,s)$ depends on the fusion rules and $C_{rs}$ is the corresponding structure constant.

Let us consider for example the correlation function of the perturbing field itself. The corresponding conformal blocks are linear combinations of products of conformal blocks at levels $1$ and $l-1$. In view of the construction (\ref{field}) there are in general 16 terms. Some of them are absent because of the fusion rules in each intermediate channel. Here there are three channels: identity $\phi_{1,1}$, the field $\tilde\phi_{1,3}$ itself and the descendent field $\tilde\phi_{1,5}$ which is defined in a way similar to that of $\tilde\phi_{1,3}$. We compute the conformal blocks up to a sufficiently high level and make a guess (remind that we need the result in the leading order in $\e\rightarrow 0$). Every internal channel should be considered separately. Let us consider for example the contribution of the identity. There is obviously a term where the identity multiplies the conformal block $F_{l-1}$ itself. A similar result comes from a product of $F_1$ with a conformal block of the field $\phi_{3,3}(l-1)$ (equal to one at this order). In addition there are "free terms" coming from some two-point functions. As a result we get a recursive equation for the conformal block $F_l$ at level $l$:
$$
F_l=a^4F_{l-1}+b^4F_1+2a^2b^2+2a^2b^2 x^2 \CC_{(13)(33)}^{(31)}(l-1)\left(1+{1\ov (1-x)^2}\right).
$$
Since the conformal block at first level (Virasoro) and the structure constant are known, this equation is easily solved. Similar considerations lead to recursive equations in the other internal channels. Combining altogether we finally obtain the 2D correlation function:
\bea\label{tpf}
&<&\tphi(x)\tphi(0)\tphi(1)\tphi(\infty)>=\\
\nn &=&\left|{1\ov x^2(1-x)^2}\left[1-2x+({5\ov 3}+{4\ov 3l})x^2-({2\ov 3}+{4\ov 3l})x^3+{1\ov 3}x^4\right]\right|^2+\\
\nn &+&{16\ov 3l^2}\left|{1\ov x(1-x)^2}\left[1-{3\ov 2}x+{l+1\ov 2}x^2-{l\ov 4}x^3\right]\right|^2+\\
\nn &+&{5\ov 9}\left({2(l-1)\ov l}\right)^2\left|{1\ov (1-x)^2}\left[1-x+{l\ov 2(l-1)}x^2\right]\right|^2.
\eea
One can check that this function is crossing symmetric and has a correct behaviour near the singular points.

We now use this function for the computation of the $\b$-function up to the second order. For that purpose we have to compute the integral in (\ref{bint}). As we explained, the above function should be integrated over the safe region. Near the singularities we use the OPE where the structure constant is known \cite{myo} and reads (up to the first order in $\e$):
$$
 C_{(1,3)(1,3)}^{(1,3)}={4\ov l\sqrt 3} - 2 \sqrt 3 \e
$$
Putting altogether we obtained that the finite part of the integral is surprisingly simple:
$$
{80\pi^2\ov 3l^2 \e^2}-{88\pi^2\ov l\e}.
$$
Taking into account also the first order term, we get the final
result (up to the second order) for the two-point function of the
perturbing field:
\bea\label{twopt}
G(x,\l)&=&<\tphi (x)\tphi(0)>\\
\nn &=&(x\bar x)^{-2+2\e}\left[1-\l {4\pi\ov \sqrt 3}\left({2\ov l\e}-3\right)(x\bar x)^\e+{\l^2\ov 2}\left({80\pi^2\ov 3l^2 \e^2}-{88\pi^2\ov l\e}\right)(x\bar x)^{2\e}
+\ldots\right].
\eea
We now introduce a renormalized coupling constant $g$ and a renormalized field $\tphi^g=\p_g {\cal L}$ analogously to $\tphi=\p_{\l}{\cal L}$.
It is normalized by $<\tphi^g (1)\tphi^g(0)>=1$.
In this renormalization scheme the $\b$-function is given by \cite{zam,pogt}:
$$
\beta(g)=\e\l{\p g \ov\p\l}=\e\l\sqrt{ G(1,\l)}
$$
One can invert this and compute the bare coupling constant and the $\beta$-function in terms of $g$:
\bea\label{bare}
\l&=&g+g^2{\pi\ov \sqrt 3}\left({2\ov l\e}-3\right)+g^3{\pi^2\ov 3}\left({4\ov l^2\e^2}-{10\ov l\e}\right)+{\cal O}(g^4),\\
\beta(g)&=&\e g-g^2{\pi\ov\sqrt 3}({2\ov l}-3\e)-{4\pi^2\ov 3l}g^3+{\cal O}(g^4).
\eea
A nontrivial IR fixed point occurs at the zero of the $\beta$-function:
\be\label{fx}
g^*={l\sqrt{3}\ov 2\pi}\e(1+{l\ov 2}\e).
\ee
It corresponds to the IR CFT  $M(k-l,l)$ as can be seen from the central charge difference:
$$
c^*-c=-{4(l+2)\ov l}\pi^2\int_0^{g^*}\beta(g)d g=-l(1+{l\ov 2})\e^3-{3l^2\ov 4}(l+2)\e^4+{\cal O}(\e^5).
$$
The anomalous dimension of the perturbing field becomes
$$
\D^*=1-\p_g\beta(g)|_{g^*}=1+\e+l\e^2+{\cal O}(\e^3)
$$
which matches with that of the field $\phi_{3,1}(l,p-l)$ of $M(k-l,l)$ (defined precisely below).

Let us define recursively the descendant fields $\tilde\phi_{n,n\pm 2}$:
\bea\nn
\tilde\phi_{n,n+2}(l,p)&=&x\phi_{n,n}(1,p)\tilde\phi_{n,n+2}(l-1,p+1)+y\phi_{n,n+2}(1,p)\phi_{n+2,n+2}(l-1,p+1),\\
\nn \tilde\phi_{n,n-2}(l,p)&=&\tilde x\phi_{n,n}(1,p)\tilde\phi_{n,n-2}(l-1,p+1)+\tilde y\phi_{n,n-2}(1,p)\phi_{n-2,n-2}(l-1,p+1)
\eea
(where $x$, $\tilde x$ and $y$, $\tilde y$ are at $(l,p)$) and the derivative $\partial\phi_{n,n}$ of the primary field
\be\nn
\phi_{n,n}(l,p)=\phi_{n,n}(1,p)\phi_{n,n}(l-1,p+1).
\ee
They have dimensions close to $1$
\bea\label{dimn}
\tilde\D_{n,n\pm 2} &=&1+{n^2-1\ov 4p}-{(2\pm n)^2-1\ov 4(p+l)}=1-{1\pm n\ov 2}\e+O(\e^2),\\
\nn 1+\D_{n,n} &=&1+{n^2-1\ov 4p}-{n^2-1\ov 4(p+l)}=1+{(n^2-1)l\ov 16}\e^2+O(\e^3).
\eea
This suggests that they mix along the RG-trajectory. To ensure this we ask that their fusion rules with the perturbing field are closed. This requirement defines the coefficients and the corresponding structure constants. So we impose the conditions:
\bea\nn
\tilde\phi_{1,3}(l,p)\tilde\phi_{n,n+2}(l,p) &=&\CC_{(13)(nn+2)}^{(nn)}(l,p)\phi_{n,n}(l,p)+\CC_{(13)(nn+2)}^{(nn+2)}(l,p)\tilde\phi_{n,n+2}(l,p),\\
\nn \phi_{3,3}(l,p)\phi_{n,n}(l,p) &=&\CC_{(33)(nn)}^{(nn+2)}(l,p)\tilde\phi_{n,n+2}(l,p)+\CC_{(33)(nn)}^{(nn)}(l,p)\phi_{n,n}(l,p).
\eea
They lead to functional equations for the coefficients and the structure constants.
In order to solve these functional equations we use the fact that we know the value of the structure constants $\CC(1,p)$, i.e. the Virasoro ones. Also, by construction, the fields $\phi_{3,3}(l,p)$ and $\phi_{n,n}(l,p)$ are primary. Finally, one can use the knowledge of the solutions for $l=1,2,4$ \cite{df,zp,pogtri}.  With all this, we can make a guess and check it directly.
Here is the solution:
\bea\nn
\CC_{(33)(nn)}^{(nn)}(l,p) &=&{\CG_n(p+l-1)\over\CG_n(p-1)},\\
\nn \CC_{(33)(nn)}^{(n+2n+2)}(l,p) &=&{\tilde\CG_n(p+l-1)\over\tilde\CG_n(p-1)},\\
\nn \CC_{(33)(nn)}^{(nn+2)}(l,p) &=&\sqrt{{l\ov (p-n-1)(p+l-n-1)}}{\tilde\CG_n(p+l-1)\over\CG_n(p-1)},\\
\nn \CC_{(33)(nn+2)}^{(n+2n+2)}(l,p) &=&-2\sqrt{{l\ov (p-n-1)(p+l-n-1)}}{\CG_{n+2}(p+l-1)\over\tilde\CG_n(p-1)}
\eea
\bea\nn
\CC_{(13)(nn)}^{(nn)}(l,p) &=&-(n-1)\sqrt{{l\ov (p+l-2)(p-2)}} \CG_n(p+l-1),\\
\nn \CC_{(13)(nn)}^{(nn+2)}(l,p) &=&\sqrt{{(p+l-2)(p-n-1)\ov (p+l-n-1)(p-2)}} \tilde\CG_n(p+l-1),\\
\nn \CC_{(13)(nn+2)}^{(nn+2)}(l,p) &=&\left(-l(n+1)+{2(p+l-2)(p-n-1)\ov p+l-n-1}\right){\CG_{n+2}(p+l-1)\ov \sqrt{l(p+l-2)(p-2)}},\\
\nn \CC_{(33)(nn+2)}^{(nn+2)}(l,p) &=&(1-{2l\ov (p-n-1)(p+l-n-1)}){\CG_{-n+2}(p+l-1)\over\CG_{-n}(p-1)}
\eea
where we introduced the functions:
\bea\nn
\CG_n(p)&=&\left[\g^3({p\ov p+1})\g^2({2\ov p+1})\g^2({n-1\ov p+1})\g^2({p-n\ov p+1})\g({3\ov p+1})\right]^{1\ov 4},\\
\nn \tilde\CG_n(p)&=&\left[\g({p\ov p+1})\g({n\ov p+1})\g({p-n-1\ov p+1})\g({3\ov p+1})\right]^{1\ov 4}.
\eea
Remind that the "structure constants" thus obtained are actually square roots of the true structure constants $C$.
The structure constants involving the field $\tilde\phi_{n,n-2}(l,p)$ are obtained from the corresponding ones for $\tilde\phi_{n,n+2}(l,p)$ by simply changing $n\rightarrow -n$.

We want to compute the matrix of anomalous dimensions and the corresponding mixing matrix of the fields defined above. For that purpose we compute their two-point functions up to second order and the corresponding integrals (\ref{bint}). The first order integrals are proportional to the structure constants we computed above. For the second order calculation we need the corresponding four point functions. They are obtained in a way similar to that of the perturbing field $\tphi(z)$ itself. Here is the list of these 4-point functions:
\bea\nn
&<&\tphi(x)\tphi(0)\tphi_{n,n+2}(1)\tphi_{n,n+2}(\infty)>=\\
\nn &=&\left|{1\ov x^2(1-x)^2}\left[1-2x+({5\ov 3}+{4\ov 3l})x^2-({2\ov 3}+{4\ov 3l})x^3+{1\ov 3}x^4\right]\right|^2+\\
\nn &+&{8\ov 3l^2}{n+3\ov n+1}\left|{1\ov x(1-x)^2}\left[1-{3\ov 2}x+{l+1\ov 2}x^2-{l\ov 4}x^3\right]\right|^2+\\
\nn &+&\left({2(l-1)\ov l}\right)^2{(n+3)(n+4)\ov 18 n(n+1)}\left|{1\ov (1-x)^2}\left(1-x+{l\ov 2(l-1)}x^2\right)\right|^2.
\eea
\bea\nn
&<&\tphi(x)\tphi(0)\tphi_{n,n+2}(1)\tphi_{n,n-2}(\infty)>=\\
\nn &=&{1\ov 3n}\sqrt{n^2-4}\left( {2(l-1)\ov l}\right)^2
\left|{1\ov (1-x)^2}\left(1-x+{l\ov 2(l-1)}x^2\right)\right|^2.
\eea
$$
<\tphi(x)\tphi(0)\tphi_{n,n}(1)\tphi_{n,n+2}(\infty)>={4\ov 3l}\sqrt{{n+2\ov n}}|x|^{-2}.
$$
\bea\nn
&<&\tphi(x)\tphi(0)\tphi_{n,n}(1)\tphi_{n,n}(\infty)>=\\
\nn &=&|x|^{-4}+{(n^2-1)\e^2\ov 12}|x|^{-4}\left({x^2\ov 2(1-x)}+{\bar x^2\ov 2(1-\bar x)}+(\log (1-x)+\log (1-\bar x))^2\right).
\eea

Let us describe briefly the renormalization scheme.
We introduce renormalized fields $\phi^g_\a$ which are expressed through the bare ones by:
\be\label{defb}
\phi^g_\a=B_{\a\b}(\l)\phi_\b
\ee
(here $\phi$ could be a primary or a descendent field).
The two-point functions of the renormalized fields
\be\label{norm}
G_{\a\b}^g(x)=<\phi_\a^g(x)\phi_\b^g(0)>,\quad G_{\a\b}^g(1)=\delta_{\a\b}
\ee
satisfy the Callan-Symanzik equation:
$$
(x\p_x-\b(g)\p_g)G_{\a\b}^g+\sum_{\rho=1}^2(\G_{\a\rho}G_{\rho\b}^g+\G_{\b\rho}G_{\a\rho}^g)=0.
$$
The matrix of anomalous dimensions $\Gamma$ that appears above is given by
\be\label{ano}
 \G=B\hat\D B^{-1}-\e\l B\p_\l B^{-1}
 \ee
where $\hat\D=diag(\D_1,\D_2)$ is a
diagonal matrix of the bare dimensions.
The matrix $B$, as defined in (\ref{defb}), is
computed from the matrix of the bare two-point functions we computed,
using the normalization condition (\ref{norm}) and requiring the matrix
$\G$ to be symmetric.

Let us combine the fields in consideration in a vector with components:
$$
\phi_1=\tphi_{n,n+2},\quad
\phi_2=(2\D_{n,n}(2\D_{n,n}+1))^{-1}\p\bar\p \phi_{n,n},\quad
\phi_3=\tphi_{n,n-2}.
$$
The field $\phi_2$ is normalized so that its bare two-point function is $1$.

We can write the matrix of the bare two-point functions $G_{\a,\b}(x,\l)=<\phi_\a(x)\phi_\b(0)>$ up to the second
order in the perturbation expansion as:
$$
G_{\a,\b}(x,\l)=
(x\bar x)^{-\D_\a-\D_\b}\left[\delta_{\a,\b}-\l C^{(1)}_{\a,\b}(x\bar x)^{\e}+{\l^2\ov 2}C^{(2)}_{\a,\b}(x\bar x)^{2\e}+...\right].
$$
As we already mentioned, the two-point functions in the first order are proportional to the
structure constants:
$$
C^{(1)}_{\a,\b}=C_{(1,3)(\a)(\b)}{\pi \g(\e+\D_\a-\D_\b)\g(\e-\D_\a+\D_\b)\ov
 \g(2\e)}.
$$
The second order contribution is a result of the double integration in (\ref{bint}) of the four-point functions we presented above. This integration goes along the same lines as in the case of the perturbing field.

Using the entries $C^{(1)}$ and $C^{(2)}$ thus obtained we can apply the renormalization procedure and obtain the matrix of anomalous dimensions (\ref{ano}). The bare coupling constant $\l$ is
expressed through $g$ by (\ref{bare}) and the bare dimensions, up to order $\e^2$. Evaluating this matrix at the fixed point (\ref{fx}), we get:
\bea\nn
\G_{1,1}^{g^*}&=&1 + {(20 - 4 n^2) \e\ov 8 (n+1)} + {l(39 - n - 7 n^2 + n^3) \e^2\ov
 16 (n+1)},\\
\nn \G_{1,2}^{g^*}&=&\G_{2,1}^{g^*}={(n-1) \sqrt{{n+2\ov n}} \e(1+l\e)\ov n+1},\\
\nn \G_{1,3}^{g^*}&=&\G_{3,1}^{g^*}=0,\\
\nn \G_{2,2}^{g^*}&=&1 + {4 \e\ov n^2-1} + {l(65 - 2 n^2 + n^4) \e^2\ov 16 (n^2-1)},\\
\nn \G_{2,3}^{g^*}&=&\G_{3,2}^{g^*}={\sqrt{{n-2\ov n}} (n+1) \e(1+l\e)\ov n-1},\\
\nn \G_{3,3}^{g^*}&=&1 + {(n^2-5) \e\ov 2 (n-1)} + {l(-39 - n + 7 n^2 + n^3) \e^2\ov
 16 (n-1)}
\eea
Its eigenvalues are (up to order $\e^2$):
\bea\nn
\D_1^{g^*}&=&1 +  {1 + n\ov 2} \e + {l(7 +8 n + n^2)\ov 16} \e^2,\\
\nn \D_2^{g^*}&=&1 + {l(n^2-1)\ov 16} \e^2,\\
\nn \D_3^{g^*}&=&1 + {1-n\ov 2} \e +  {l(7 - 8 n + n^2)\ov 16} \e^2.
\eea
This result coincides with the dimensions $\tilde\D_{n+2,n}(l,p-l)$, $\D_{n,n}(l,p-l)+1$ and $\tilde\D_{n-2,n}(l,p-l)$ of the model $M(k-l,l)$ up to this order.
The corresponding normalized eigenvectors should be identified with the fields of $M(k-l,l)$:
\bea\nn
\tphi_{n+2,n}(l,p-l)&=&{2 \ov n (n+1)}\phi_1^{g^*} + {2
\sqrt{{n+2\ov n}}\ov n+1}\phi_2^{g^*} + {\sqrt{n^2-4}\ov
n}\phi_3^{g^*},\\
\nn \phi_2(l,p-l)&=&-{2 \sqrt{{n+2\ov n}}\ov n +
1}\phi_1^{g^*} -{n^2-5\ov n^2+1}\phi_2^{g^*} +{2\sqrt{{n-2\ov n}}\ov n-1}\phi_3^{g^*},\\
\nn \tphi_{n-2,n}(l,p-l)&=&{\sqrt{n^2-4}\ov n}\phi_1^{g^*}  - { 2
\sqrt{{n-2\ov n}}\ov n-1}\phi_2^{g^*} +{ 2\ov n(n-1)}\phi_3^{g^*}.
\eea
We used as before the notation $\tphi$ for the descendent field defined as in the UV theory and:
$$
\phi_2(l,p-l)={1\ov 2\D_{n,n}^{p-l}(2\D_{n,n}^{p-l}+1)}\p\bar\p
\phi_{n,n}(l,p-l)
$$
is the normalized derivative of the corresponding primary field. We notice that these eigenvectors are finite
as $\e\rightarrow 0$ with exactly the same entries as in $l=1$ \cite{pogt} and  $l=2$ \cite{as} minimal models. This is one of the main results of this paper.


\begin{thebibliography}{99}

\bibitem{gko}{P. Goddard, A. Kent, D. Olive, Virasoro algebras and coset space models, \pl{152}{1985}88.}
%
\bibitem{pogt}{R. Poghossian, Two Dimensional Renormalization Group Flows in Next to Leading Order, JHEP {\bf 1401} (2014) 167; \hepth{1303.3015}.}
%
\bibitem{as}{C. Ahn, M. Stanishkov, On the Renormalization Group Flow in Two Dimensional Superconformal Models, \np{885}{2014}713; \hepth{1404.7628}.}
%
\bibitem{myo}{C. Crnkovic, G. Sotkov, M. Stanishkov, Renormalization group flow for general $SU(2)$ coset models, \pl{226}{1989}297.}
%
\bibitem{myt}{C. Crnkovic, R. Paunov, G. Sotkov, M. Stanishkov, Fusions of Conformal Models, \np{336}{1990}637.}
%
\bibitem{kmq}{D. Kastor, E. Martinec, Z. Qiu, Current Algebra and Conformal Discrete Series, \pl{200}{1988}434.}
%
\bibitem{zam}{A. Zamolodchikov, Renormalization Group and Perturbation Theory Near Fixed Points in Two Dimensional Field Theory, Sov. J. Nucl. Phys. {\bf 46} (1987) 1090.}
%
\bibitem{df}{V. Dotsenko, V. Fateev, Operator Algebra of Two Dimensional Conformal Theories with Central  Charge $c<1$, \pl{154}{1985}291}
%
\bibitem{zp}{A. Zamolodchikov, R. Poghossian, Operator algebra in two dimensional superconformal field theory, Sov. J. Nucl. Phys. {\bf 47}(1988)929.}
%
\bibitem{pogtri}{R. Poghossian, Operator Algebra in Two Dimensional Conformal Quantum Field Theory Containing Spin $4/3$ Parafermionic Coserved Currents, \ijmp{A6}{1991}2005.}
%


\end{thebibliography}
\end{document}